\documentclass[10pt,twoside,twocolumn,english,aps,prl,showpacs]{revtex4-1}
\pdfoutput=1
\usepackage[latin9]{inputenc}

\usepackage{wasysym}
\usepackage{graphicx}
\usepackage{esint}
\usepackage[unicode=true, pdfusetitle,
 bookmarks=true,bookmarksnumbered=false,bookmarksopen=false,
 breaklinks=false,pdfborder={0 0 1},backref=false,colorlinks=false]
 {hyperref}
\usepackage{color,soul}

\makeatletter

\newcommand{\binom}[2]{{#1 \choose #2}}

\@ifundefined{textcolor}{}
{%
 \definecolor{BLACK}{gray}{0}
 \definecolor{WHITE}{gray}{1}
 \definecolor{RED}{rgb}{1,0,0}
 \definecolor{GREEN}{rgb}{0,1,0}
 \definecolor{BLUE}{rgb}{0,0,1}
 \definecolor{CYAN}{cmyk}{1,0,0,0}
 \definecolor{MAGENTA}{cmyk}{0,1,0,0}
 \definecolor{YELLOW}{cmyk}{0,0,1,0}
 }

\renewcommand{\fnum@figure}{Fig.~\thefigure}

\makeatother

\begin{document}

\title{Stroboscopic observation of quantum many-body dynamics}

\author{Stefan Ke{\ss}ler,$^{1}$ Andreas Holzner,$^{2}$ Ian P. McCulloch,$^{3}$
Jan von Delft,$^{2}$ and Florian Marquardt$^{1,4}$}

\affiliation{$^{1}$Institute for Theoretical Physics, Universit\"{a}t Erlangen-N\"{u}rnberg,
Staudtstr. 7, DE-91058 Erlangen, Germany\\$^{2}$Physics Department, Arnold Sommerfeld Center for Theoretical
Physics, and Center for NanoScience, Ludwig-Maximilians-Universit\"{a}t
M\"{u}nchen, Theresienstr. 37, DE-80333 M\"{u}nchen, Germany\\$^{3}$School of Physical Sciences, The University of Queensland,
Brisbane, Queensland 4072, Australia\\$^{4}$Max Planck Institute for the Science of Light, G\"{u}nther-Scharowsky-Straße
1/Bau 24, DE-91058 Erlangen, Germany}

\begin{abstract}
Recent experiments have demonstrated single-site resolved observation
of cold atoms in optical lattices. Thus, in the future it may be possible
to take repeated snapshots of an interacting quantum many-body system
during the course of its evolution. Here we address the impact of
the resulting quantum (anti-)Zeno physics on the many-body dynamics.
We use the time-dependent density-matrix renormalization group to obtain the
time evolution of the full wave function, which is then periodically projected
in order to simulate realizations of stroboscopic measurements. For the 
example of a one-dimensional lattice of spinless fermions with nearest-neighbor
interactions, we find regimes for which many-particle configurations
are stabilized or destabilized, depending on the interaction strength
and the time between observations. 
\end{abstract}


\maketitle
\emph{Introduction.} In the last few years ultracold atoms in optical
lattices have proven to be a versatile tool for studying various quantum
many-body phenomena \citep{Bloch2008,Lewenstein2007}. Recently, tremendous
progress has been achieved by implementing single-site resolved detection
\citep{Bakr2009,Sherson2010} and addressing \citep{Weitenberg2011}
of atoms. Taken to the next, dynamic level, one may envision observing the 
evolution of nonequilibrium quantum many-body states via periodic snapshots 
revealing the position of each single atom. For simpler systems, the effect 
of frequent observations on the decay of an unstable state (or on the dynamics 
of a driven transition) has already been discussed and observed, leading
to the notion of the quantum (anti-)Zeno effect \citep{Misra1977,Itano1990,
Fischer2001,Facchi2008}. Zeno physics has also been seen in cold-atom experiments 
with atomic loss channels \citep{Syassen2008} and was theoretically addressed
in \citep{Garcia-Ripoll2008,Daley2009,Schuetzhold2010,Shchesnovich2010}. 
Experiments with single-site detection, however, would reveal the effect
of observations on the dynamics of a truly interacting quantum many-body
system. Here we exploit a numerically efficient approach to simulating
the repeated observation of many-particle configurations in interacting
lattice models. This represents an idealized version of the dynamics
that may be realized in future experiments. We elaborate the main features of 
this ``stroboscopic'' many-body dynamics in the case of a one-dimensional (1D) 
lattice of  spin-polarized fermions with nearest-neighbor interactions. We find 
a variant of the quantum Zeno effect and discuss its tendency to inhibit or 
accelerate the break-up of certain many-particle configurations. In particular, 
the decay rate of these configurations depends in a \emph{nonmonotonous} fashion 
on the time interval between observations. Later, we show that a similar behavior 
is expected for the Fermi-Hubbard and Bose-Hubbard model. The discussed features 
may be seen, e.\,g., in the expansion of interacting atomic clouds in a lattice.

\emph{Model.} In this paper, we study spin-polarized fermions in
a 1D lattice governed by the Hamiltonian
\begin{equation}
\hat{\mathcal{H}}=-J\sum_{i}\big(\hat{c}_{i}^{\dagger}\hat{c}_{i+1}+\mbox{H.c.}\big)+V\sum_{i}\hat{n}_{i}\hat{n}_{i+1}\,.\label{eq: hamiltonian}
\end{equation}
The first term describes hopping with amplitude $J$ between adjacent sites, the 
second encodes the interaction between fermions at neighboring sites, with 
$\hat{n}_{i}=\hat{c}_{i}^{\dagger}\hat{c}_{i}$. The Hamiltonian displays a dynamical
$V\mapsto-V$ symmetry which shows up in expansion experiments \citep{Schneider2010}.
Following analogous steps as in \citep{Schneider2010}, we can conclude that if both 
the initial state and the experimentally measured quantity $\hat{\mathcal{O}}$ are 
invariant under both time reversal and $\pi$--boost (a translation of all momenta 
by $\pi$), the observed time evolution $\langle\hat{\mathcal{O}}(t)\rangle$ is 
identical for repulsive and attractive interactions of the same strength. The initial 
occupation number states and the $n$-particle density observables in our case fall 
within the scope of this theorem. Thus, the only relevant dimensionless parameters are
$|V/J|$ and the rescaled time between observations, $J\Delta t$. 

\begin{figure}
\includegraphics[scale=0.19]{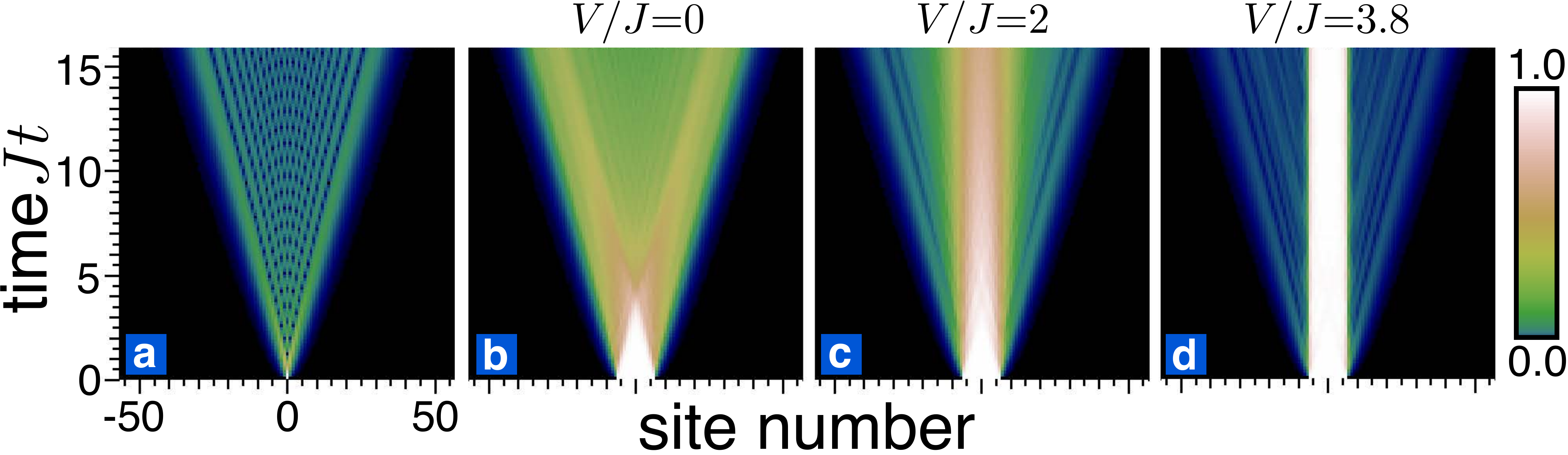}
\caption{(Color online) Time evolution of the density profile of fermions expanding in a lattice
without observation, (a) for a single fermion; (b)-(d) for 
13 fermions, initially located at adjacent lattice sites, calculated for several
interaction strengths. \label{fig:expansion_density_profile}}
\end{figure}

\emph{Single particle.} We first briefly turn to the single-particle case, in which 
Eq.~(\ref{eq: hamiltonian}) leads to a tight-binding band $E(k)=-2J\cos(k)$. A particle 
located initially at a single site is in a superposition of all plane wave momenta 
$k=-\pi\ldots\pi$. After a time $t$, the probability of detecting it at a distance $l$
from the initial site is $\rho(l,t)=\mathcal{J}_{|l|}^{2}(2Jt)$, where $\mathcal{J}$ is 
the Bessel function of the first kind. This is shown in 
Fig.~\ref{fig:expansion_density_profile}(a). The particle moves ballistically, with 
$\langle l^{2}\rangle=2(Jt)^{2}$. When the particle is observed repeatedly, at intervals 
$\Delta t$, the ballistic motion turns into diffusion. In this case, after $m$ time steps 
of duration $\Delta t=t/m$, we have $\langle l^{2}\rangle=2J^{2}t\,\Delta t$. Thus the 
motion slows down, and in the limit of an infinite observation rate, the particle is frozen, 
which is known as the quantum Zeno effect. 

\emph{Simulation.} Ideally, each observation is a projective measurement in the basis of 
many-particle configurations (occupation number states in real space). To generate such 
measurement outcomes in a numerically efficient way, we start by randomly drawing the 
position of the first particle from a distribution given by the one-particle density.
Afterward, we draw the position of the second particle, \emph{conditioned} on the location
of the first one, and proceed iteratively (see Supplemental Material \citep{Supplement} 
for details). For the interacting many-body case to be discussed now, we use the 
time-dependent density-matrix renormalization group (tDMRG) \citep{Vidal2004,Daley2004,
White2004,Schmitteckert2004} to calculate the joint probabilities and the time evolution 
between observations. We choose a time step of $J\delta t=0.1$ and a lattice of typically 
115 sites, and we keep up to approximately $1000$ states, at a truncation error of $10^{-6}$. 
For the dynamics of noninteracting fermions, an exact formula can be used (see Supplemental 
Material \citep{Supplement}). Note that tDMRG has been employed recently for dissipative
dynamics of cold atoms \citep{Pichler2010, Barmettler2011}.

\emph{Stroboscopic many-body dynamics.} We will focus on the expansion of an interacting 
cloud from an initially confined state. For 2-species fermions, such an expansion has been 
recently observed in 2D \citep{Schneider2010} and theoretically discussed for 1D 
\citep{Kajala2011}. We will first briefly address the evaporation itself and then discuss
qualitatively the resulting stroboscopic dynamics, with a more refined analysis presented 
further below. Figures~\ref{fig:expansion_density_profile}(b)-(d) show the effect of the 
interaction on the unmeasured time evolution of the density profile. For increasing 
interaction the fermions tend to remain localized near their initial positions. For 
interaction strengths $|V/J|\apprge3$ and the times shown here, $Jt<16$, a more detailed 
analysis reveals that evaporation proceeds via the rare event of a single fermion 
dissociating from the edge of the cloud. The particle then moves away ballistically. This 
evaporation process is hindered by the formation of bound states. This is a crucial 
phenomenon that we will also encounter in the context of repeated measurements. For smaller 
interaction strengths ($|V/J|\apprle2$), the fermions split gradually into a larger and 
larger number of clusters as time increases. The parameter regimes in which the model 
(\ref{eq: hamiltonian}) exhibits diffusive or ballistic transport was addressed using tDMRG
in Ref. \citep{Langer2009}.
\begin{figure}
\includegraphics[scale=0.19]{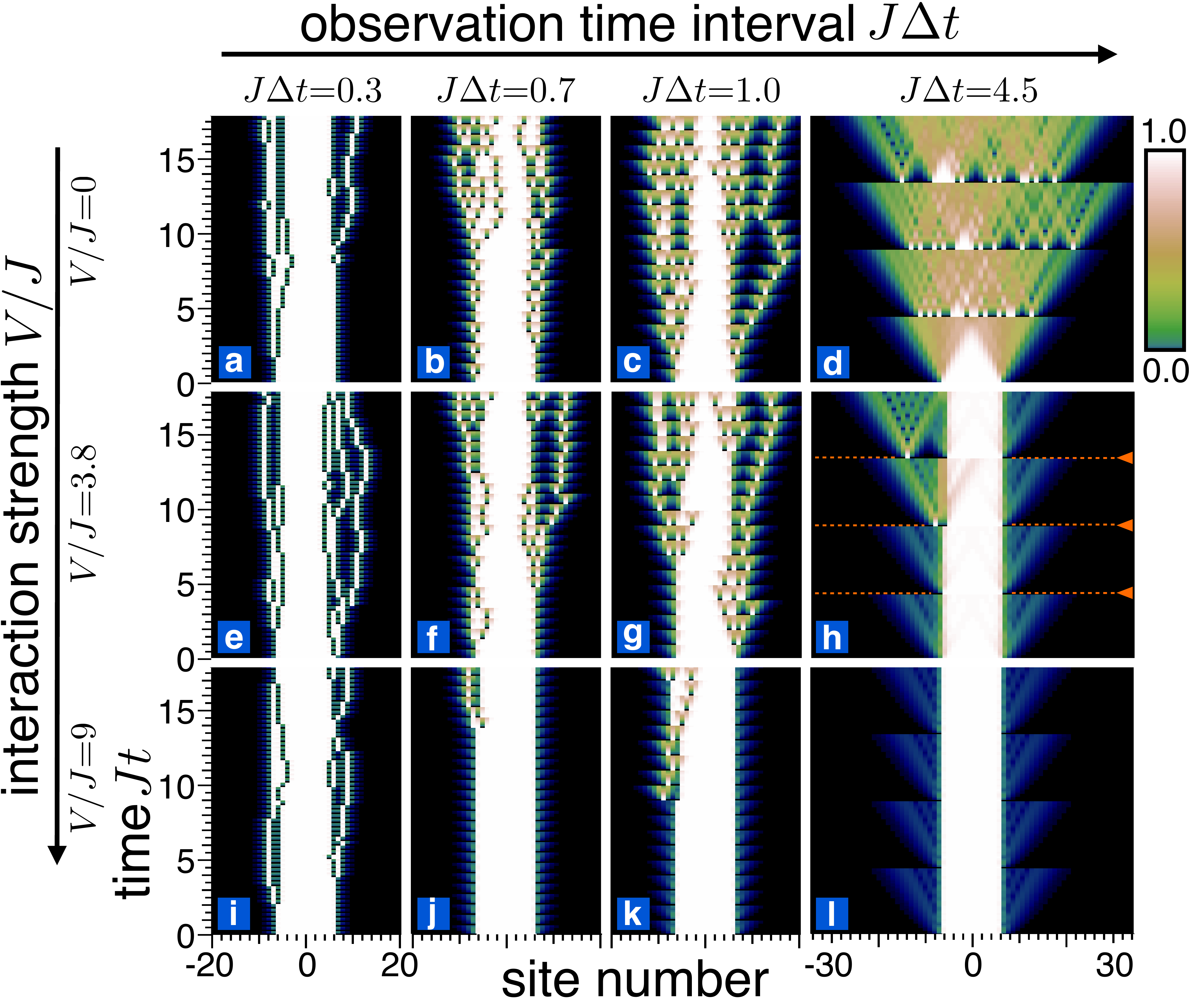}
\caption{(Color online) Specific realizations of the expansion of 13 fermions
with site-resolved detection during the evolution. We show the full
evolution of density even between observations, which collapse the
many-body wave function at regular time intervals $J\Delta t$ [indicated by the dashed lines 
in panel (h)]. Without interaction [panels (a)-(d)], the initial decay rate of the 
configuration increases for larger $J\Delta t$, while for large interaction $|V/J|$ the decay 
rate is biggest for finite $J\Delta t$ [(g),(i)]. For small $J\Delta t$ the dynamics becomes 
independent of $\left|V/J\right|$, see (a),(e),(i). \label{fig:stroboscopic_expansion}}
\end{figure}

\begin{figure}
\includegraphics[scale=0.188]{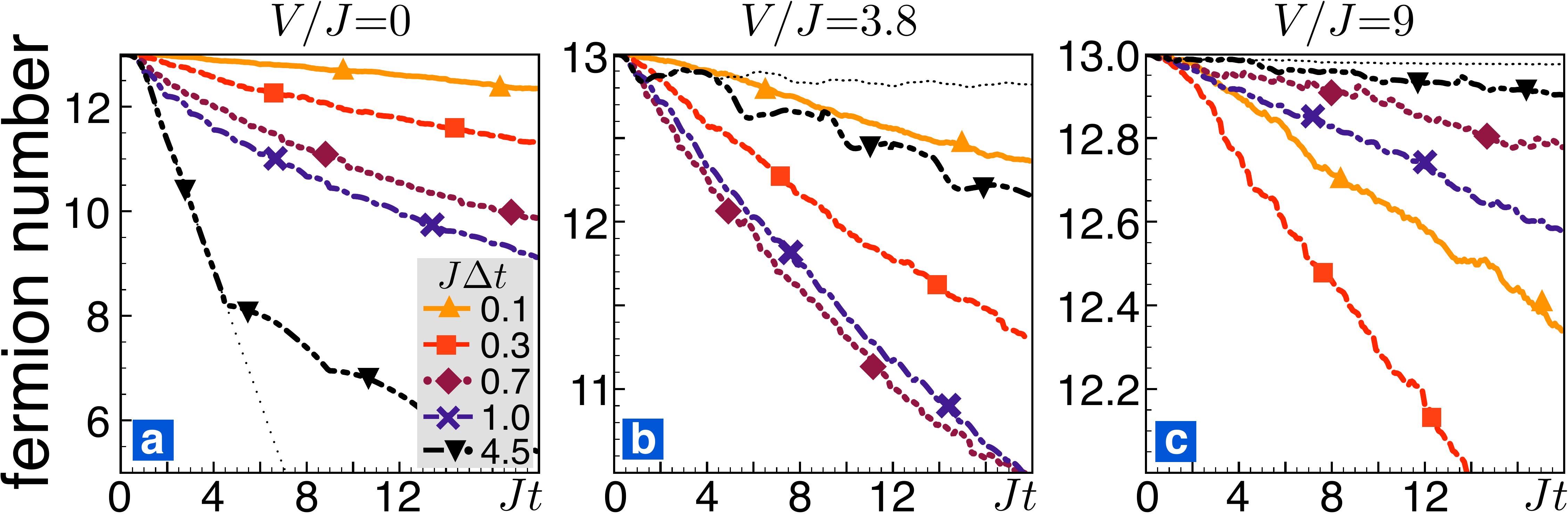}
\caption{(Color online) For the expansion process of Fig.~\ref{fig:stroboscopic_expansion}, the 
average number of fermions remaining at the 15 central lattice sites is shown as function of the 
evolution time $Jt$ for different observation time intervals $J\Delta t$. Thin dotted lines 
correspond to expansions without observation. (a) Without interactions, the initial decay rate 
increases monotonically with $J\Delta t$. (b) At $V/J=3.8$, the decay rate first increases, then 
decreases with $J\Delta t$. (c) At $V/J=9$, the decay rate is nonmonotonous even for intermediate 
$J\Delta t$; compare $J\Delta t=0.3,0.7,1.0$. Note that the lines in (a)-(c) are almost identical 
for $J\Delta t=0.1$. \label{fig:nonmonotonic_behavior}}
\end{figure}

The effects of stroboscopic observation are shown in Fig.~\ref{fig:stroboscopic_expansion},
for typical realizations of this stochastic process. For noninteracting fermions we find the 
behavior expected from the single-particle case. The spread (and thus, the diffusion constant) 
increases with larger observation time intervals $J\Delta t$. For very small $J\Delta t$
(strong Zeno effect), the motion is diffusive with a small diffusion constant that becomes 
independent of $\left|V/J\right|$. In general, it is useful to discuss the 
\emph{initial decay rate} of the cluster that evaporates via expansion. For the interacting case, 
this decay rate is largest at some finite observation time interval $J\Delta t$ 
[Figs.~\ref{fig:stroboscopic_expansion}(g) and \ref{fig:stroboscopic_expansion}(i)], while it 
is reduced for large $J\Delta t$ [Figs.~\ref{fig:stroboscopic_expansion}(h) and 
\ref{fig:stroboscopic_expansion}(l)]. Apparently, at very large $\left|V/J\right|$, the initial 
decay rate may have yet another local minimum for intermediate $J\Delta t$, see 
Fig.~\ref{fig:stroboscopic_expansion}(j). We confirm this striking nonmonotonous behavior of 
the initial decay rate by simulating 400 realizations for each panel shown in 
Fig.~\ref{fig:stroboscopic_expansion} and plot the average number of fermions at the central 
15 lattice sites as a function of time, in Fig.~\ref{fig:nonmonotonic_behavior}.
For sufficiently large $\left|V/J\right|$, this number decays roughly linearly at a rate that 
sets the initial decay rate. We will see that these features can be mainly attributed to a bound 
state and the two-level dynamics between the initial state and the state with a fermion detached 
from the others. 

\emph{Doublets and the role of interactions.} The effect of interactions can be discussed already 
for the stroboscopic dynamics of two fermions. We focus on the decay of a doublet, i.\,e., two 
fermions sitting at neighboring sites.
 
In the quantum Zeno limit, $J\Delta t\ll$1 (or $J\Delta t\ll\left|2J/V\right|$ for large 
$\left|V\right|$, see below), only single hopping events occur during $\Delta t$. The probability 
for a fermion hopping left or right in this time interval is $2(J\Delta t)^{2}$. Thus, the average 
decay time of a doublet is $\langle Jt\rangle=1/(2J\Delta t)$, independent of $V$. This result 
holds also for clusters of more fermions, where the leftmost and rightmost fermion dissolves with 
probability $(J\Delta t)^{2}$ during $\Delta t$ [cf. Figs.~\ref{fig:stroboscopic_expansion}(a),
\ref{fig:stroboscopic_expansion}(e), and \ref{fig:stroboscopic_expansion}(i)].

\begin{figure}
\includegraphics[scale=0.19]{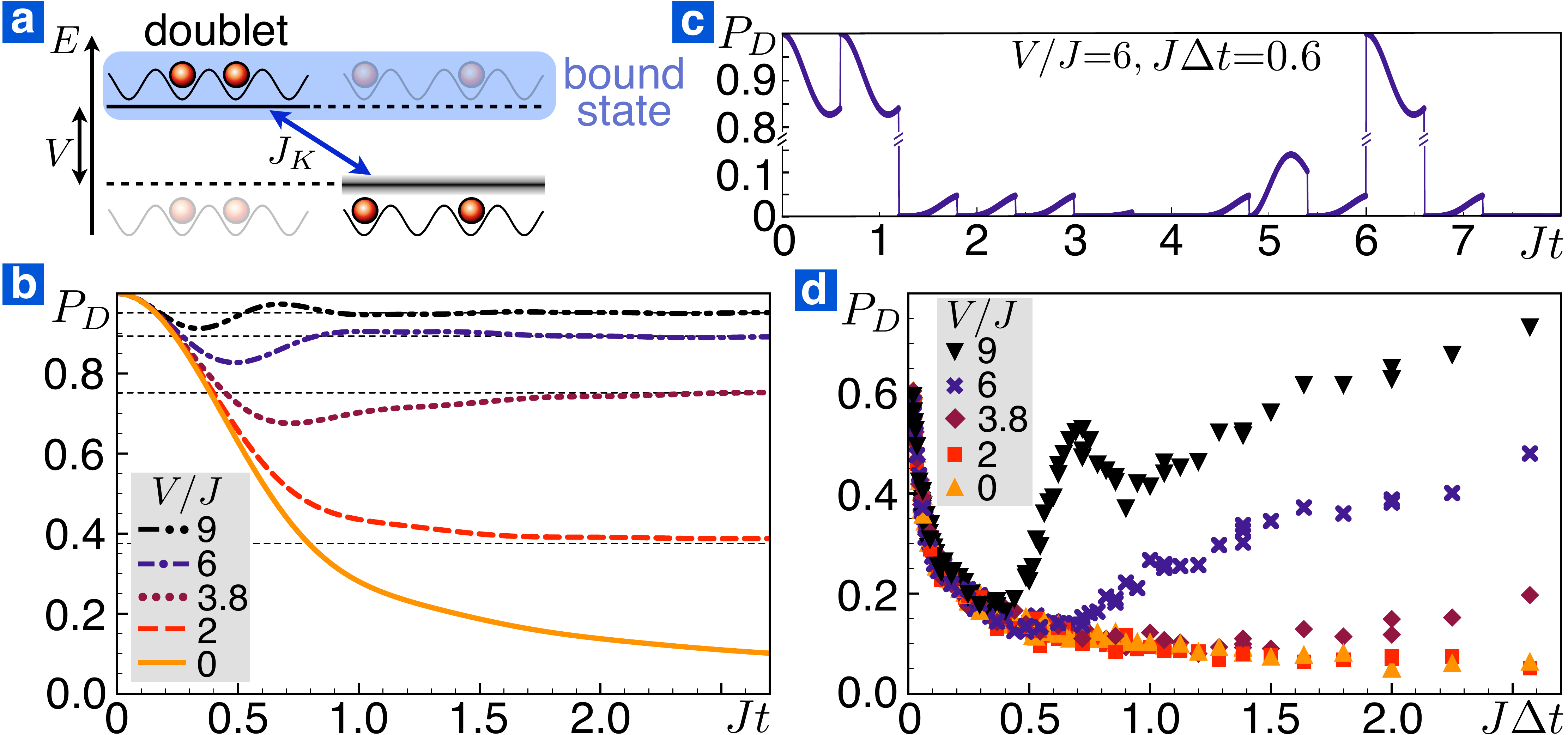}
\caption{(Color online) (a) Doublet decay level scheme. The doublet is separated from the
continuum of unbound states by an energy gap $V$. (b) Probability
$P_{D}$ of finding the doublet intact after evolution time $Jt$.
Dashed horizontal lines show $P_{D}(\infty)$ found in Eq.~(\ref{eq:doublet_probability_infinite_time}).
(c) Single trajectory of $P_{D}$ for a time evolution subject to observations. (d) Doublet survival
probability $P_{D}$ as function of the observation time interval
$J\Delta t\geq0.02$ for a fixed total evolution time $Jt=18$. Note
the non-monotonous dependence on $J\Delta t$ for finite interactions.
\label{fig:doublet_decay}}
\end{figure}

For larger $J\Delta t$, the interaction will become important, which gives rise to a bound state 
(this effect also exists for clusters of more particles \citep{Sutherland2004}). Considering the 
basis $|l,K\rangle=\frac{1}{\sqrt{\mathcal{N}}}\sum_{j}\exp\{iK[j+l/2]\}c_{j}^{\dagger}c_{j+l}^{\dagger}|\textrm{vac}\rangle$
of the 2-particle sector with (positive) relative coordinate $l$, center-of-mass (c.m.)
coordinate $j+l/2$, and total wavenumber $K=(k_{1}+k_{2})\,\textrm{mod}\,2\pi$, the action of the 
Hamiltonian (\ref{eq: hamiltonian}) is 
$\hat{\mathcal{H}}|l,K\rangle=|K\rangle\otimes\hat{\mathcal{H}}_{K}|l\rangle$.
The first part describes a plane wave with wavenumber $K$, the second the relative motion given by
\begin{equation}
\hat{\mathcal{H}}_{K}|l\rangle=-2J_{K}\big[|l+1\rangle+(1-\delta_{l,1})|l-1\rangle\big]+V\delta_{l,1}|l\rangle,\label{eq:effective hamiltonian}
\end{equation}
with $K$-dependent hopping amplitude $J_{K}=J\cos(K/2)$. A bound state exists if $|V|\geq|2J_{K}|$. 
It is given by $|\psi_{K}\rangle\propto\sum_{l=1}^{\infty}(-2J_{K}/V)^{l-1}|l\rangle$. We now discuss 
the decay of a doublet [see Fig.~\ref{fig:doublet_decay}(a)] by means of the doublet survival
probability $P_D(t)$, i.e., the probability of finding the doublet intact after time $t$. Without 
observations, $P_D(t)=\sum_{L'} |\langle l=1,L'|e^{-i\hat{\mathcal{H}}t} |l=1,L\rangle|^2$, where $L$ 
and $L'$ are c.m. coordinates. Thus, in the limit $t\rightarrow\infty$, we find
\textbf{ $P_{D}(\infty)=\frac{1}{2\pi}\int_{0}^{2\pi}\mathrm{d}K\left|\langle\psi_{K}|l=1\rangle\right|^{4}$}.
Specifically for large interaction $|V/2J|\geq1$, we have
\begin{equation}
P_{D}(\infty)=1-\left(2J/V\right)^{2}+\frac{3}{8}\left(2J/V\right)^{4}\,.\label{eq:doublet_probability_infinite_time}
\end{equation}
While $P_{D}(\infty)$ is determined by the bound state, the evolution for times $Jt<1$ can be 
approximated by the two-level dynamics between $|l=1\rangle$ and $|l=2\rangle$. This gives
\begin{equation}
P_{D}(t)=1-\frac{1}{\pi}\int_{0}^{\pi}\mathrm{d}K\frac{\cos^{2}(K/2)}{\xi_{K}^{2}}\sin^{2}(2\xi_{K}Jt),\label{eq:doublet_decay_short_time}
\end{equation}
with $\xi_{K}=\left[(\frac{V}{4J})^{2}+\cos^{2}(K/2)\right]^{1/2}$.In the strongly interacting 
regime we find three regions for the doublet survival probability: for times $Jt\ll\xi_{K=0}^{-1}$
the probability is independent of the interaction strength, $P_{D}(t)=1-2(Jt)^{2}$; for times 
$\xi_{K=0}^{-1}\apprle Jt\apprle 1$ one expects an oscillating behavior of $P_{D}(t)$ given by 
Eq.~(\ref{eq:doublet_decay_short_time}) with a period approximately $\frac{2\pi}{V}$ for 
$|V/4J|\gg1$; and for $Jt\gg1$ the probability approaches $P_{D}(\infty)$.
The full evolution of $P_{D}(t)$ using exact diagonalization is shown in 
Fig.~\ref{fig:doublet_decay}(b). $P_{D}(t)$ is interaction-independent at times 
$Jt\apprle\xi_{K=0}^{-1}/2$. Temporal oscillations in $P_{D}(t)$ develop for higher interaction 
strengths ($V/J\apprge3.5$). These oscillations suggest that in the presence of stroboscopic 
observations, illustrated in Fig.~\ref{fig:doublet_decay}(c), the survival probability will depend
nonmonotonically on the observation time interval. This effect is confirmed in 
Fig.~\ref{fig:doublet_decay}(d). In that figure, the observation time interval $J\Delta t$ is varied, 
while keeping the total evolution time constant, $Jt=18$ (with a corresponding number of observations 
$t/\Delta t$). The stroboscopic evolution is interaction independent for small $J\Delta t$. For larger 
$J\Delta t$ there is a drastic recovery of $P_{D}$ in the strong interacting case, which can show 
oscillations as a function of $J\Delta t$. This behavior agrees with the one of clusters of more 
fermions, see Figs.~\ref{fig:stroboscopic_expansion} and \ref{fig:nonmonotonic_behavior}, and does not 
depend in detail on the total time $Jt$. Thus we have found and explained the most prominent features 
of the stroboscopic many-body dynamics in our discussion of the doublet.

\begin{figure}
\includegraphics[scale=0.23]{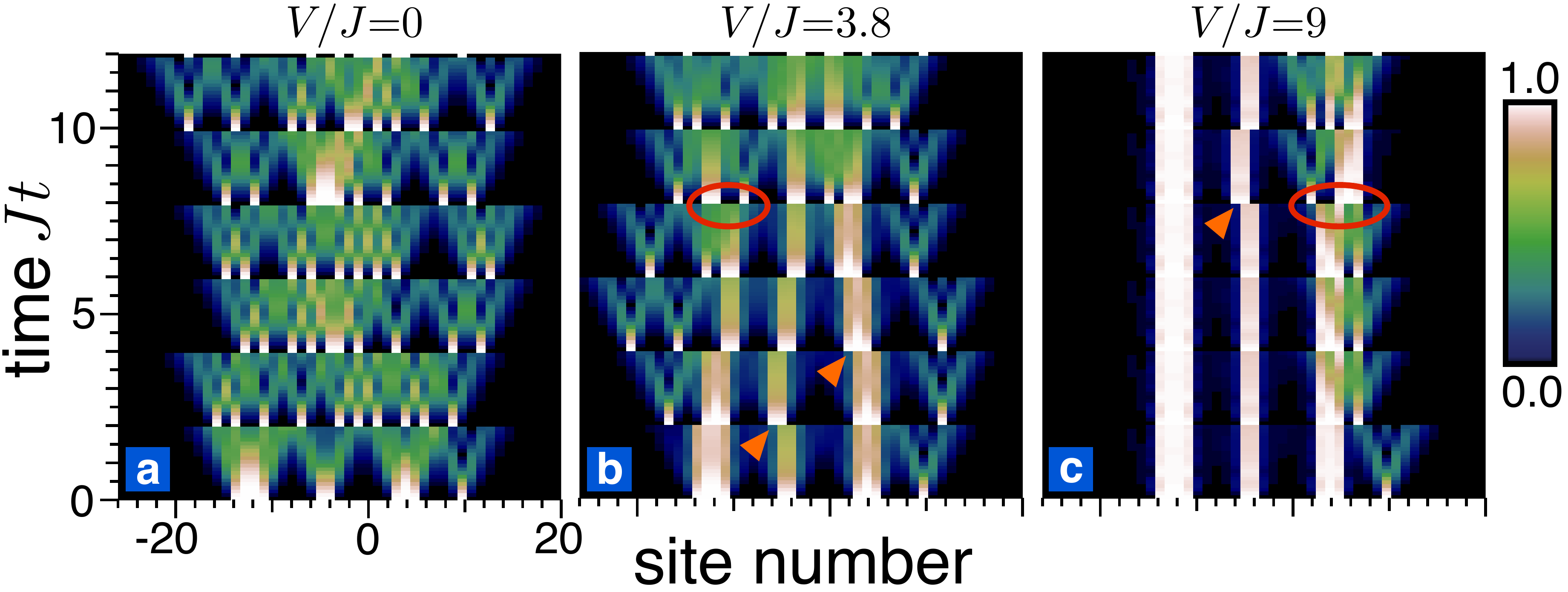}
\caption{(Color online) Density plot for the time evolution of an initial state with clusters of different 
numbers of fermions and observation time interval $J\Delta t=2$. For large interaction strength 
[panels (b) and (c)], we find clusters moving as a whole (indicated by triangles). Ovals indicate processes 
where single fermions are exchanged between clusters or attached to a new cluster.}
\label{fig:tracks_of_disorder}
\end{figure}

Furthermore, the motion of whole clusters of fermions through the lattice and the exchange of fermions 
between clusters can be observed in the stroboscopic dynamics, as shown in Fig.~\ref{fig:tracks_of_disorder}.
As expected, clusters are very stable for high interaction strengths. The hopping amplitude for a 
cluster of $n$ fermions is of order $J^{n}/\left|V\right|^{n-1}$, decreasing strongly for larger clusters, 
as can be perceived in Fig.~\ref{fig:tracks_of_disorder}(c).

\begin{figure}
\includegraphics[scale=0.2]{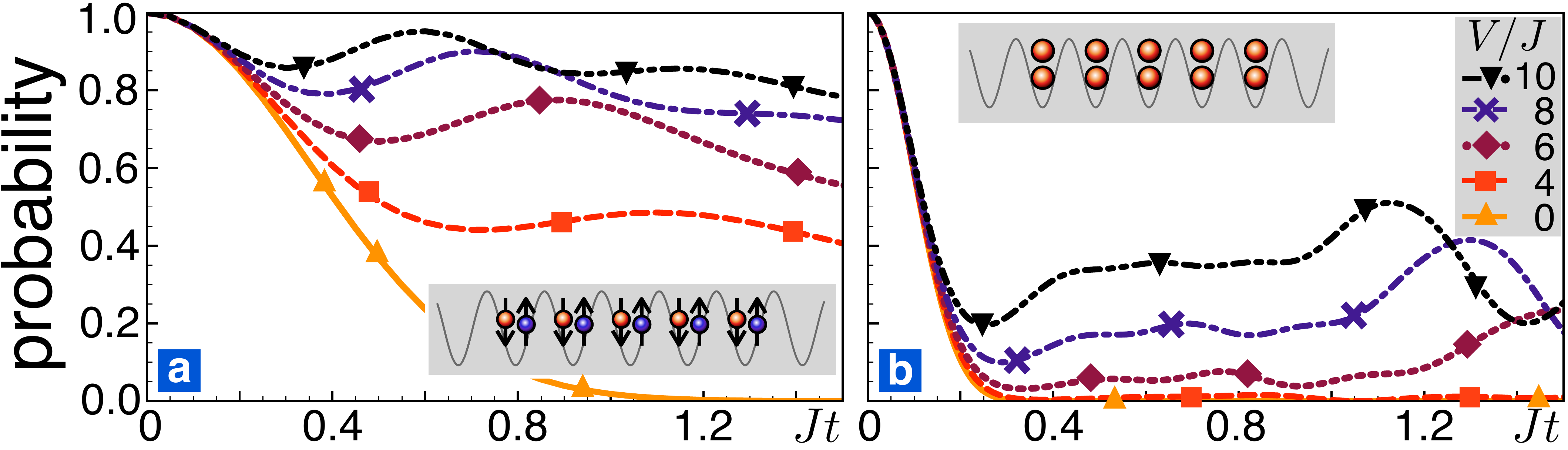}
\caption{(Color online) Destruction of a cluster in the 1D Fermi-Hubbard (a) and Bose-Hubbard model (b). 
We show the probability to detect a cluster of five doublons, see insets, in its initial state as a
function of time for different onsite interaction strengths $V$.\label{fig:decay_FH_BH_clusters}}
\end{figure}

\emph{Expectations for other models.} The previously discussed nonmonotonic decay of a cluster is not 
unique for our model (\ref{eq: hamiltonian}), but can also be observed in the Fermi-Hubbard (FH) and 
Bose-Hubbard (BH) models. In the following we will consider clusters of doubly occupied sites in 1D.
The dissolution of a single particle from the edge of the cluster is well described by considering 
the survival probability of a doublon (fermionic or bosonic double occupancy). Using a similar analysis
as for the doublets, one finds qualitatively the same behavior as shown in Fig.~\ref{fig:doublet_decay}(b). 
However, this description is not sufficient to understand the decay of clusters. It does not include the 
escape of paired particles, which leads to a complete decay of the cluster in the long time limit even 
for large onsite interaction. It also neglects tunneling of bosons between occupied sites.
In the FH model the cluster is initially Pauli blocked [as in model (\ref{eq: hamiltonian})], such that 
the dynamics is restricted to the edges. The escape of single fermions or doublons can be well 
approximated by the Hubbard dimer model, see \citep{Kajala2011}. The associated probability  is 
$\sin^{2}(\sqrt{V^{2}+(4J)^{2}}t/2)/[2+2(J/4V)^{2}]$ for single fermions (at times $Jt\ll1$) and for 
doublons it is  $(Jt)^{4}$ for $Jt\ll J/V$ and $4(J^{2}t/V)^{2}$ at $J/V\ll Jt\ll1$. We numerically 
verified that this oscillating behavior of the single fermion escape leads to nonmonotonic decay of the 
cluster for onsite interaction strengths $V/J\apprge4$, see Fig.~\ref{fig:decay_FH_BH_clusters}(a).
The probability for an initial bosonic cluster configuration of $N$ sites to be destroyed is for short
times given by $4[1+3(N-1)](Jt)^{2}$. The first term in brackets stems from a single boson escaping from 
the edge, while the second and larger contribution comes from a boson tunneling between occupied sites. 
For larger times we studied numerically the destruction of clusters 
[see Fig.~\ref{fig:decay_FH_BH_clusters}(b)] and find that the survival probability of the initial cluster 
configuration displays a nonmonotonic behavior for $V/J\apprge6$ with a pronounced local minimum at 
$t\approx\pi/V$ (about half the oscillation period for the dissolution of a boson and tunneling of a boson 
between doubly occupied sites), and the cluster is destroyed primarily via ``internal'' hopping processes. 
Both models show a strong nonmonotonic behavior of the cluster decay probability for large onsite 
interaction strengths. Thus, by tuning the time interval between observations in a stroboscopic measurement, 
one is able to enhance and suppress the decay of clusters, or even different decay channels of the cluster.

\emph{Experimental realization and outlook.} The Hamiltonian (\ref{eq: hamiltonian}) is related to the 
Heisenberg $XXZ$ model by Wigner-Jordan transformation. The stroboscopic dynamics is identical for both 
models as the outcome of observations depends only on spatial density-density correlations. These 
Hamiltonians can be experimentally realized in optical lattices with fermionic polar molecules \citep{Buchler2007}
or 2-species fermions or bosons in the insulating phase \citep{Duan2003,Garcia-Ripoll2003}. For both 
realizations single-site detection has not yet been implemented, but experimental progress is being made 
toward this goal. Experimentally, the most challenging step needed to observe the interplay of many-body
dynamics and measurements discussed here would be to make the observations nondestructive, whereas currently 
atoms are heated into higher site orbitals and atom pairs are lost due to light-induced collisions 
\citep{Bakr2009,Sherson2010}. Beyond the scenarios discussed here, one may also be interested in the influence 
of external driving or measurements that either are weak or target only specific sites.
\\

Financial support by the DFG through NIM, SFB/TR 12, and the Emmy-Noether program is gratefully acknowledged. 

\bibliographystyle{apsrev4-1}
\bibliography{Stroboscopic}

\clearpage{}

\appendix
\onecolumngrid

\section*{Supplemental Material}

\subsection*{Numerical implementation of the single-site resolved observation}

An ideal single-site resolved observation of particles is a projective measurement 
in the basis of many-particle configurations (occupation number states in real space). 
However, due to the exponentially large number of states, we need a numerically efficient 
way to sample such outcomes. This is achieved by generating the measurement outcomes in a 
stepwise fashion, building on the fact that (positive) $n$-particle densities $\rho_{n}$ 
factorize into conditional probabilities
\begin{equation}
\rho_{n}(s_{1},\ldots,s_{n})=\rho_{1}(s_{1})\cdot\prod_{i=2}^{n}\rho_{i}(s_{i}|s_{i-1},\ldots,s_{1}).
\end{equation}
Here, $s_{i}\in\{1,\ldots,N_{s}\}$ denotes the site of the $i$th particle and 
$\rho_{i}(s_{i}|s_{i-1},\ldots,s_{1})$ is the conditional probability of finding the $i$th 
particle at site $s_{i}$ given that there are $i-1$ particles at the sites $s_{1},\ldots,s_{i-1}$.
The procedure starts by randomly drawing the position of the first particle from a distribution 
given by the one-particle density. In the next step, we draw the position of the second 
particle, \emph{conditioned} on the location of the first one, and continue iteratively. This 
way less than $n\cdot N_{s}$ values of joint probability densities have to be calculated, in 
comparison to the full number $\binom{N_{s}}{n}$ of fermionic many-body configurations. This 
approach relies on being able to calculate efficiently both the pure time evolution between
observations and the $i$-particle densities ($1\leq i\leq n$). In the present work, we use the 
time-dependent density-matrix renormalization group, which is an extremely powerful method for 
interacting one-dimensional systems. For the model considered in the main article, it is 
numerically even more efficient to draw the position of the first particle as before and then 
project the state onto those configurations where a particle is present at the selected site. 
After rescaling the resulting state, the new one-particle density is calculated. From this 
distribution we draw the position of the second fermion, excluding all sites already occupied by 
a fermion,and iterate the steps for the remaining fermions. For a discussion about local 
measurements on quantum many-body systems using matrix product states, see \cite{Gammelmark2010}.

\subsection*{Stroboscopic dynamics of noninteracting fermions}

In the noninteracting case, we can explicitly calculate the $n$-particle densities needed to 
simulate the stroboscopic dynamics. After each observation, the many-particle wave function is a 
Slater determinant of single-particle wave functions, and for noninteracting fermions this remains 
true even during the subsequent time evolution. Using Wick's theorem, the $n$-particle density of 
$N$ fermions at time $t$ can be written as (see also \cite{Lowdin1955}):
\begin{equation}
\rho_{n}(s_{1},\ldots,s_{n};t)=\left|\begin{array}{ccc}
M_{s_{1}s_{1}}(t) & \ldots & M_{s_{1}s_{n}}(t)\\
\ldots &  & \ldots\\
M_{s_{n}s_{1}}(t) & \ldots & M_{s_{n}s_{n}}(t)\end{array}\right|.\end{equation}
Here, $M_{s_{k}s_{l}}(t)=\sum_{j=1}^{N}T_{s_{k}m_{j}}(t)T_{s_{l}m_{j}}^{*}(t)$, with the fermion 
propagator $T_{kl}(t)=i^{(k-l)}\mathcal{J}_{k-l}(2Jt)$, and the sum is taken over all initially 
occupied sites $m_{j}$. Note that the one-particle density $\rho_{1}(t)$ is just the sum of the
densities of the individual fermions, cf. Fig.1(b) in the main article. Motion in arbitrary 
potentials would be captured by different propagators $T_{kl}(t)$.

\end{document}